# Efficient Channel Model for Homogeneous Weakly Coupled Multicore Fiber

Lin Gan, Jiajun Zhou, Songnian Fu, Ming Tang, *Senior Member, IEEE* and Deming Liu

***Abstract*—To analyze a homogeneous weakly coupled multicore fiber (WC-MCF) based transmission system via simulation, we propose an efficient (fast and accurate) WC-MCF's channel model, which can describe the propagation effects including attenuation, walk-off, chromatic dispersion, self-phase modulation (SPM), and especially the frequency-dependent inter-core crosstalk (XT). We speed up the simulation with two orders of magnitude by simplifying the XT's calculation. On one hand, the calculation step size can be greatly increased by utilizing a new XT's coupling matrix. On the other hand, the calculation of XT can be further accelerated by down-sampling XT's coupling matrix in frequency domain. The XT power and average occurrence distance should be set manually based on the existing XT model to describe the frequency-dependent XT the same as a real WC-MCF. We numerically and experimentally observed that XT's de-correlation bandwidth decreases with relative time delay (RTD) by fractional linear function. The range of validity of the proposed channel model is also discussed with different walk-off and coupling strength. We believe the proposed efficient channel model can provide great help for analysis and optimization of homogeneous WC-MCF based optical communication systems.**

## I. Introduction

As a kind of realization of space division multiplexing (SDM) technology, multicore fiber (MCF) has been widely used in various communication scenarios in recent years to enhance transmission capacity [1]-[3]. MCF can be roughly classified into weakly-coupled multicore fiber (WC-MCF) [4]-[9] and strongly-coupled multicore fiber (SC-MCF) [10]-[14]. SC-MCF is more advantageous in long-haul transmission scenarios because of its better nonlinear tolerance. However, the crosstalk (XT) in SC-MCF between different cores must be mitigated by complex digital signal processing algorithms, such as multiple-input and multiple-output (MIMO) techniques [14]. On the contrary, due to low XT in WC-MCF, it is not necessary to mitigate the XT impacts via complex MIMO. Therefore, for cost-sensitive scenarios, such as inter-data center optical interconnect [15], optical access network [16], [17] and front haul for 5G [18], [19], WC-MCF is more competitive compared with SC-MCF.

To optimize the WC-MCF based transmission systems, one of the most meaningful issues is to investigate the XT penalty. Based on this, the WC-MCF can be carefully designed to meet the maximum acceptable threshold of XT. Therefore, several experiments have been conducted in recent years. For example, in [6], the authors concluded that 1-dB signal-to-noise (SNR) penalty at a bit-error ratio of $10^{-3}$ from XT is -18 dB, -24 dB, and -32 dB for QPSK, 16-QAM, and 64-QAM, respectively. In [20], the authors investigated the performance fluctuation of an adaptive DD-OFDM MCF link affected by XT. In order not to be limited to specific experimental conditions, several WC-MCF's channel models have been proposed to analyze the XT impacts on transmission systems [14], [21]-[26].

These channel models are always based on the nonlinear Schrodinger equations (NLSEs) with coupling terms added to describe XT. Usually, there are two strategies to solve the NLSEs. The first strategy is to use small calculation step size (such as 1.0 μm to 1.0 mm) to solve NLSEs as accurately as possible based on the equivalent refractive index (ERI) model [14], [21], [22]. Based on this strategy, the evolution of XT at each phase-matching point (PMP) can be well described. Therefore, this strategy has high calculation accuracy. But it is too time consuming to cover the transmission distance with tens of kilometers or even longer. Moreover, the calculation step size cannot be set larger which is limited by the accurate calculation of the ERI model. The second strategy is to use large calculation step size (such as 100.0 m to 1.0 km), but it needs to ignore some characteristics of XT to simplify calculation and decrease simulation time [23]-[25]. Usually, XT is calculated only once in each calculation step size (such as 100.0 m) modeled as signal copies with random phase shift. Namely, it is assumed that there is only one PMP within a large calculation step size. However, the second strategy cannot truly describe the nature of XT in time domain or in frequency domain, which will be further proved in this paper. Thus, it is necessary and meaningful to build an efficient (fast and accurate) channel model for WC-MCF which describes XT accurately and has small computational effort.

It should be noted that to describe XT accurately in WC-MCF' channel model, the XT related parameters should

Manuscript received November 22, 2018. This work was supported in part by National Natural Science Foundation of China under Grant (61722108, 61331010, 61205063, 61290311); Major Program of the Technical Innovation of Hubei Province of China (2016AAA014); the Open Fund of State Key Laboratory of Optical Fiber and Cable Manufacture Technology, YOFC (SKLD1706) and the Fundamental Research Funds for the Central Universities': HUST: 2018KFYXKJC023. *(Corresponding author: Ming Tang.)*

L. Gan, J. Zhou, S. Fu, M. Tang, and D. Liu are with Wuhan National Lab for Optoelectronics (WNLO) & National Engineering Laboratory for Next Generation Internet Access System (NGIA), School of Optics and Electronic Information, Huazhong University of Science and Technology (HUST), Wuhan 430074, China (e-mail: lingan@hust.edu.cn; jiajunzhou@hust.edu.cn; songnian@mail.hust.edu.cn; tangming@mail.hust.edu.cn; dmliu@mail.hust.edu.cn). L. Gan and J. Zhou contribute equally to this work.

first be calculated from the existing XT models, such as the XT average power with specific fiber properties (core refractive index profile, core pitch) or under specific external conditions (bending, twisting) [27]-[31]. It also should be noted that the time-dependent XT in each simulation has always not been considered, because the decorrelation time of the short term average crosstalk (STAXT) is about several minutes which is much higher than a typical time window of simulation (2.3 microseconds for $2^{16}$ symbols with 28 Gbaud) [30].

In our previous work [26], the transmission performance of 100 Gbps PDM-QPSK signals was numerically investigated based on a simple MCF's channel model (Fig. 2 in [26]). However, the de-correlation bandwidth needs to be set manually. Therefore, in [8] and [9], we observed the relationship between de-correlation bandwidth and relative time delay. Further, based on it, we built a new WC-MCF's channel model. But the calculation step size cannot be higher than several meters to describe XT at each PMP, which means it is too time consuming.

In this paper, we propose an efficient (fast and accurate) channel model for homogeneous WC-MCF, which describes the frequency-dependent crosstalk (XT) almost the same as a real WC-MCF. Firstly, detailed description about the proposed WC-MCF's channel model is introduced. Secondly, the channel model is validated via experimental results. Finally, the scope of validation of the proposed WC-MCF's channel model is discussed with different coupling strength. We believe the proposed efficient channel model can provide great help for analysis and optimization for homogeneous WC-MCF based optical communication systems.

## II. Efficient Channel Model for Homogeneous WC-MCF

It can be assumed that there are only two cores (Core A and Core B) in homogeneous WC-MCF. The XT caused by multiple interfering cores will not be discussed in this paper to focus on the principle description of the proposed channel model. When there is no XT between Core A and Core B, the optical signals in each core can be well described by the NLSE shown as (1), without the rightest term which describes the coupling between cores [32]. $z$ and $t$ represent longitudinal coordinate and time, respectively. There are two individual polarization states ($x$ and $y$) for both Core A and Core B, in which the electric field component can be represented as $E_{x,y}^{A,B}$. These polarization components correspond to the principal axes of polarization in each fiber section. For example, $E_y^B$ represents the electric field component of polarization state $y$ in Core B. The NLSE contains a differential operator $\hat{D}$ and a nonlinear operator $\hat{N}$. The differential operator $\hat{D}$ describes the propagation effects represented as (2), such as polarization mode dispersion $\beta_{1x,y}$ (PMD), walk-off $\Delta v_g^{A,B}$, chromatic dispersion $\beta_2$ (CD), dispersion slope $\beta_3$ and attenuation $\alpha$. It should be noted that the PMD is modeled by coarse-step method which is applied in [26], [33]. The values of $\beta_{1,x}$ and $\beta_{1,y}$ of current scattering section $z_{scatt,i}$ are defined as (3). And $z_{scatt}$ is randomly selected from a Gaussian distribution with mean and standard deviation set by the parameters mean step size $\mu_{scatt}$ and step size deviation $\sigma_{scatt}$ shown as (4). The walk-off describes differential group velocity between Core A and Core B [34]. It will cause small relative time delay (RTD) between cores after transmission, which can be also referred to as the inter-core skew [7] represented as (5), where $L$ represents the transmission distance. $v_g^{A,B}$ represents the group velocity in a specific core when PMD is not considered. At last, the nonlinear operator $\hat{N}$ can be represented as (6) assuming complete polarization mixing, in which $\gamma$ is the nonlinearity coefficient [32].

$$\frac{\partial E_{x,y}^{A,B}(z,t)}{\partial z} = \left(\hat{D}+\hat{N}\right)E_{x,y}^{A,B}(z,t) - j\kappa\exp\left(-j\Delta\beta^{A,B}z\right)E_{x,y}^{B,A}(z,t) \quad (1)$$

$$\hat{D} = -\left(\beta_{1x,y}+\frac{1}{\Delta v_g^{A,B}}\right)\frac{\partial}{\partial t} - \frac{j\beta_2}{2}\frac{\partial^2}{\partial t^2} + \frac{\beta_3}{6}\frac{\partial^3}{\partial t^3} - \frac{\alpha}{2} \quad (2)$$

$$\beta_{1,x,i} = -\beta_{1,y,i} = \frac{PMD}{2\sqrt{z_{scatt,i}}} \quad (3)$$

$$f(z_{scatt}) = \frac{1}{\sqrt{2\pi}\sigma_{scatt}}\exp\left(\frac{-(z_{scatt}-\mu_{scatt})^2}{2\sigma_{scatt}^2}\right) \quad (4)$$

$$\Delta v_g^A = -\Delta v_g^B = \frac{v_g^B - v_g^A}{2} = \frac{1}{2}\frac{RTD}{L} \quad (5)$$

$$\hat{N} = j\gamma\frac{8}{9}\left(\left|E_{x,y}^{A,B}(z,t)\right|^2 + \left|E_{y,x}^{A,B}(z,t)\right|^2\right) \quad (6)$$

Based on the modified coupled mode theory (CMT) for MCF with step-index profile, it has been proved that, no matter the XT is weak or strong, the transmission effects can be modeled as (1) with the rightest term added [35]. However, the (5) in [35] is only focused on the continuous wave with single carrier frequency. When there are multiple optical frequencies, the mismatch of propagation constant between cores can be extended as (7). The accumulated phase mismatch $\Delta\beta z$ of different optical frequencies are different, which can be caused by walk-off and the difference of chromatic dispersion in different cores. $\omega_0$ and $\Delta\omega$ represent reference optical radian frequency and radian frequency deviation, respectively.

$$\Delta\beta^A = -\Delta\beta^B = \left(\beta_0^B-\beta_0^A\right) + \left(\beta_1^B-\beta_1^A\right)\Delta\omega + \frac{\left(\beta_2^B-\beta_2^A\right)}{2}\Delta\omega^2 + \frac{\left(\beta_3^B-\beta_3^A\right)}{6}\Delta\omega^3 \quad (7)$$

As mentioned above, there are two strategies to solve (1). The first strategy is to choose small calculation step size (such as 1.0 μm to 1.0 mm) [14], [30], [31]. However, for a real transmission system, the fiber length could reach several kilometers to hundreds of kilometers. Therefore, it is very time consuming if we choose small calculation step size. The second strategy is to choose large calculation step size (such as 100.0 m to 1000.0 m). Usually, the split-step Fourier method is chosen to accelerate calculation shown as (8) [27], [28]. $h$ is the calculation step size, and the total fiber length can be represented as $L = Mh$.

$$E(z,t) \approx \exp\left(-\frac{1}{2}h\hat{N}\right)\cdot\left[\prod_{m=1}^{M}\exp\left(h\hat{N}\right)\exp\left(h\hat{D}\right)\right] \cdot\exp\left(\frac{1}{2}h\hat{N}\right)E(0,t) \tag{8}$$

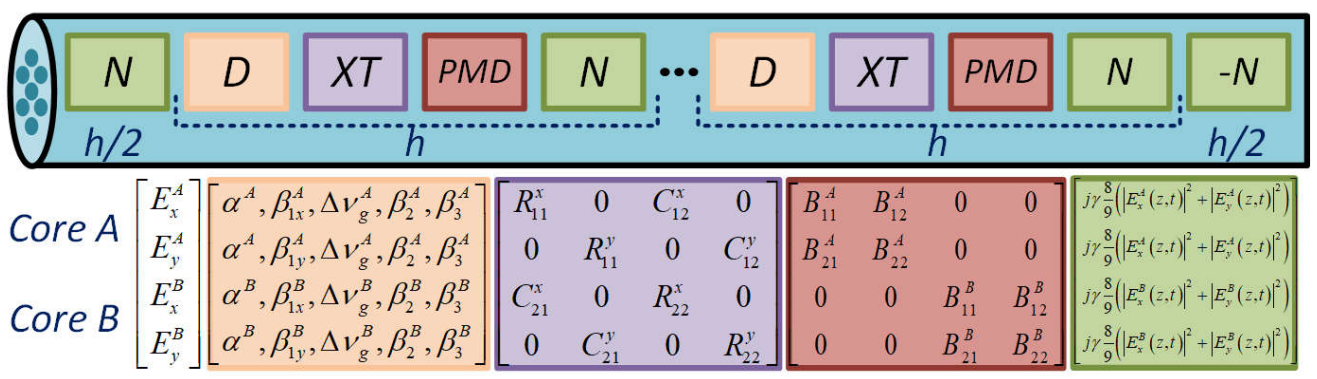

Fig. 1. Solution process for homogeneous WC-MCF's channel model.

Fig. 1 demonstrates the solution process of the proposed homogeneous WC-MCF's channel model with two cores taking into consideration. Firstly, the linear propagation effects of two electric field components in both Core A and Core B are calculated independently in each calculation step size $h$ as shown as the first color block of the bottom of Fig. 1. For example, the first row describes the attenuation $\alpha^A$, the difference group velocity between polarization states $\beta_{1x}^A$, the walk-off $\Delta v_g^A$, chromatic dispersion $\beta_2$ and dispersion slope $\beta_3$ for polarization state $x$ in Core A. We assume the attenuation, walk-off, chromatic dispersion, and dispersion slope are the same for both polarization states. However, it also can be set independently if necessary.

Secondly, a coupling matrix is used to model the XT between cores shown as the second color block. The XT is assumed only occurs in the same polarization state between Core A and Core B in each calculation step size, and two polarizations' XT are independent with each other because of slight difference of propagation constant between different polarization states.

Thirdly, the polarization random coupling is calculated using another rotation matrix based on the coarse-step method which is applied in [21], [33] shown as the third color block. For example, (9) shows the rotation matrix for random coupling of polarizations in Core A where $\theta$ is the random rotation of axes and $\varphi$ is the random phase difference between two orthogonal polarization electric fields. $E(z,\omega)$ is the Fourier-transform of optical signal $E(z,t)$, where $\omega$ is the frequency in the Fourier domain. Therefore, $\Delta\omega$ in (7) is equal to $\omega-\omega_0$. We assume that the polarization random coupling between cores are independent with each other. At last, the nonlinear effects in each core can be conducted assuming there is full mixing between different polarization states.

$$\begin{bmatrix} E_x^A(z',\omega) \\ E_y^A(z',\omega) \end{bmatrix} = \begin{bmatrix} B_{11}^A & B_{12}^A \\ B_{21}^A & B_{22}^A \end{bmatrix} \begin{bmatrix} E_x^A(z,\omega) \\ E_y^A(z,\omega) \end{bmatrix} = \begin{bmatrix} \cos\theta & e^{j\varphi}\sin\theta \\ -e^{-j\varphi}\sin\theta & \cos\theta \end{bmatrix} \begin{bmatrix} E_x^A(z,\omega) \\ E_y^A(z,\omega) \end{bmatrix} \tag{9}$$

It should be noted that the calculation of the differential operator $\hat{D}$ and the nonlinear operator $\hat{N}$ are conducted in frequency-domain and time-domain, respectively [36]. The calculation of XT and random coupling between two polarizations are conducted in frequency-domain [8]. The calculation of polarization random coupling needs to be arranged after the calculation of XT to avoid phase change induced by polarization random coupling.

In [14], after transmission within one small calculation step size of 1.0 mm, the accumulated XT in one step size can be calculated using (10). Here, taking the electric field component $E_x$ as an example, $R_{11} \approx R_{22} \approx 1$ and $R_{12} \approx 0$ are for negligible XT, and $R_{12}$ is a real value ranging from 0.0 to 1.0 depending on the XT strength. However, the coupling matrix has two deficiencies. The first is the total energy of output optical pulses after XT calculation is not rigorous equal to that of input optical pulses. The second is the calculation step size cannot be set larger (such as 100.0 m) due to the calculation of the ERI model, which means it is too time consuming for long distance simulation (such as 10.0 km to 100.0 km).

$$\begin{bmatrix} E_x^A(z',\omega) \\ E_x^B(z',\omega) \end{bmatrix} = \begin{bmatrix} R_{11} & R_{12} \\ -R_{12} & R_{22} \end{bmatrix} \begin{bmatrix} E_x^A(z,\omega) \\ E_x^B(z,\omega) \end{bmatrix} \tag{10}$$

Therefore, the problem is how to describe XT accurately with only once calculation within a large calculation step size. To solve this problem, we introduce a similar coupling matrix as (11). The main diagonal elements $R_{11}$ and $R_{12}$ in (10) are retained with real values. It means the power coupling from Core A to Core B is assumed to only affect the electric field amplitude in Core A without affecting its phase due to low XT in WC-MCF. However, the values of $R_{11}$ and $R_{12}$ cannot be set 1.0 to ensure rigorous conservation of energy after XT calculation.

$$\begin{bmatrix} E_x^A(z',\omega) \\ E_x^B(z',\omega) \end{bmatrix} = \begin{bmatrix} R_{11}(\omega) & C_{12}(\omega) \\ C_{21}(\omega) & R_{22}(\omega) \end{bmatrix} \begin{bmatrix} E_x^A(z,\omega) \\ E_x^B(z,\omega) \end{bmatrix} \tag{11}$$

On the other hand, it has been verified that the XT ($R_{12}$ in (10)) should be a complex value in a real homogeneous WC-MCF after several PMPs [4], [5]. In [34], the XT expression at longitudinal coordinate $z_k$ has been obtained as (12). $K$ represents mean coupling strength which has been discussed in [5]. $\Delta\phi_x(z+h_k,\omega)$ represents the phase mismatch in one calculation step size induced by propagation constant mismatch, which can be obtained by using (13) based on the phase transfer function (PTF) proposed in [8].

$$C_{12}(\omega) = -jK\sum_{k=1}^{N}\exp\left(-j\Delta\phi_x\left(z+h_k,\omega\right)\right) \tag{12}$$

$$\Delta\phi_x(z+h_k,\omega) = \frac{h_k}{h}\left[\phi_x^B(z,\omega)-\phi_x^A(z,\omega)\right] +\phi_x^{rnd}(z+h_k) \tag{13}$$

The PTF of electric field component $E_x$ in Core A and Core B can be represented as (14). The *angle* means the electrical field phase of each optical frequency.

$$\phi_x^{A,B}(z,\omega) = angle\left(E_x^{A,B}(z+h,\omega)\right) -angle\left(E_x^{A,B}(z,\omega)\right) \tag{14}$$

Based on the modified coupled mode theory and the ERI model [4], [5], when the homogeneous WC-MCF is twisted with 0.5 round/m, there should exist about 100 PMPs in one calculation step size $h$ of 100.0 m. Therefore, it can be assumed that the average occurrence distance of XT $h_{xt}$ is 1.0 m. The subscript $k$ in (12) ranges from 1 to $N$, which can be obtained by $N = h/h_{xt}$ . For each $h_k$ , it can be randomly selected within [0,$h$] according to the uniform distribution. $\phi_x^{rnd}\left(z+h_k\right)$ is also assumed uniformly distributed within [0,2 $\pi$ ] which is caused by the mismatch of $\beta_0$ in (7) with random twisting and bending. If necessary, to model the time-dependent XT, each random phase could be assumed stationary and modeled by independent Brownian motions [30].

After the calculation of $C_{12}$ using (12)-(14), the values of $C_{21}$ , $R_{11}$ and $R_{22}$ can be solved as (15), (16), respectively, based on the conservation of energy for arbitrary input signals represented as (17). The *conj* represents complex conjugation function.

$$C_{21}\left(\omega\right) = -conj\left(C_{12}\left(\omega\right)\right) \quad (15)$$

$$R_{11}\left(\omega\right) = R_{22}\left(\omega\right) = \sqrt{1-\left|C_{12}\left(\omega\right)\right|^2} \quad (16)$$

$$\left|E_x^A\left(z',\omega\right)\right|^2 + \left|E_x^B\left(z',\omega\right)\right|^2 = \left|E_x^A\left(z,\omega\right)\right|^2 + \left|E_x^B\left(z,\omega\right)\right|^2 \quad (17)$$

As shown in Table 1, we assume that the two cores in homogeneous WC-MCF have the identical attenuation, CD, dispersion slope and PMD level. An optical pulse can be injected into Core A with polarization state $x$ . The bandwidth of optical pulse is 50.0 GHz with up-sampled by 16 times [9]. The target XT power is set manually by solving the coupling strength based on [5]. Because the time window of the simulation is about 163.8 ns (symbol number/symbol rate), which is far smaller than the decorrelation time (few minutes) [37], it is reasonable to assume that the frequency-dependent XT remains constant in the time window. In addition, the performance fluctuation caused by time-dependent XT could be investigated by running the simulation with numerous times.

TABLE I
PARAMETERS USED IN SIMULATION

| Parameters | Values |
|---|---|
| Reference optical frequency | 193.12 THz |
| Symbol rate | 50.0 GBaud |
| Symbol number | $2^{13}$ |
| Sample per symbol | $2^4$ |
| Fiber length $L$ | 20.0 km |
| Calculation step size $h$ | 100.0 m |
| Attenuation $\alpha$ | 0.23 dB/km |
| Walk-off | $1.0\times10^{-12}$ s/m |
| Chromatic dispersion CD | $16.7\times10^{-16}$ s/m$^2$ |
| Dispersion slope | $0.08\times10^{3}$ s/m$^3$ |
| Nonlinear index $n_2$ | $2.6\times10^{-20}$ $m^2/W$ |
| Effective core area | $80.0\times10^{-12}$ m$^2$ |
| Polarization mode dispersion PMD | $0.1\times10^{-12}$ $s/\sqrt{km}$ |
| Mean step size of PMD $\mu_{scatt}$ | 100.0 m |
| Step size deviation of PMD $\sigma_{scatt}$ | 10.0 m |
| Target XT power $XT_\mu$ | -11.1 dB/100 km |
| Average occurrence distance of XT $h_{xt}$ | 1.0 m |
| Down-sampling ratio of XT $R_{xt}$ | 64 |

Based on CPU Intel i5-6500 and MATLAB 2017a, the simulation was repeated 10 times to get the average simulation time. When the WC-MCF's channel model in [9] is applied, the maximum calculation step size $h$ can only be set to 1.0 m with twisting speed of 0.5 round/m. As shown in Fig. 2(a), the average simulation time is 2559.3 seconds. However, when the proposed new WC-MCF's channel model is applied, the maximum calculation step size $h$ can be set to 100.0 m or larger. As shown in Fig. 2(b), the average simulation time is reduced significantly to 147.8 seconds due to large calculation step size of 100.0 m without down-sampling. However, we can also observe that the calculation of XT ((12)-(14)) takes about 93% (137.3 s/147.8 s) of the simulation time based on the proposed new channel model. Thus it is necessary to continually reduce the simulation time of XT's calculation.

Focused on this purpose, a new strategy to accelerate calculation is proposed, which is to down-sample the XT's coupling matrix (represented as (11)) in frequency-domain with down-sampling ratio $R_{xt}$ . The details of the strategy are as followed. After the calculation of the differential operator $\hat{D}$ in each calculation step, the PTF of two electric field components in two cores can be obtained by using (14). Because the signals are now described in frequency-domain, the PTF can be directly down-sampled in frequency-domain with down-sampling ratio $R_{xt}$ . Then, $C_{12}\left(\omega\right)$ can be obtained by using (13) and (14) based on down-sampled PTF. Lastly, $C_{12}\left(\omega\right)$ needs to be up-sampled by spline method [36] with up-sampling ratio $1/R_{xt}$ to guarantee the calculation accuracy.

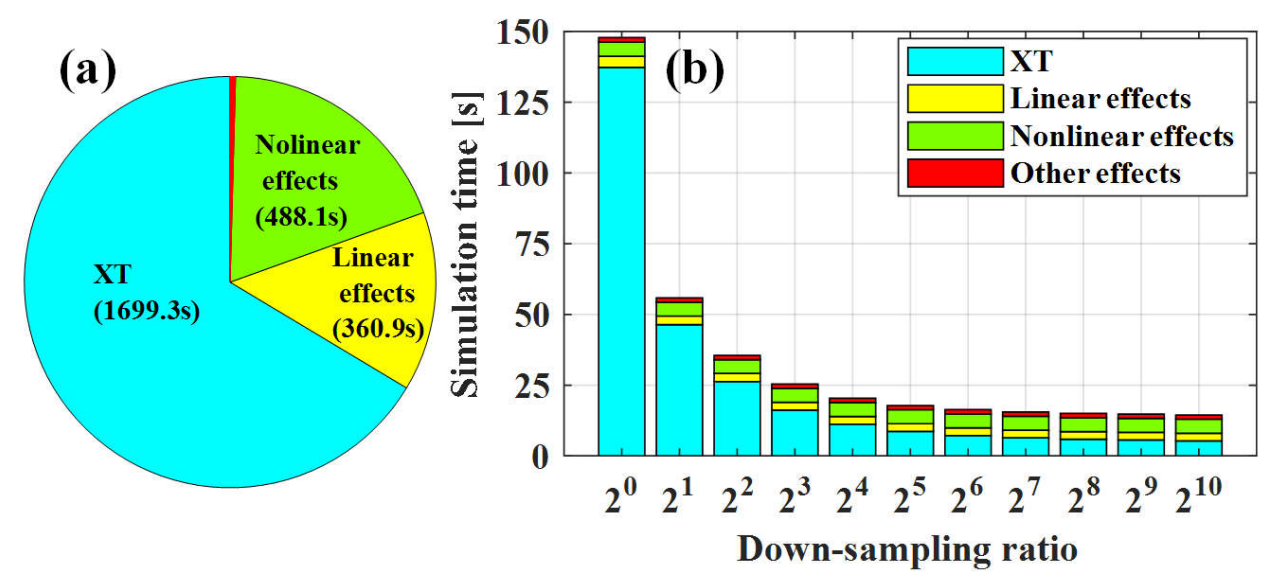

Fig. 2. (a) Simulation time with calculation step size of 1.0 m. (b) Simulation time of different down-sampling ratios with calculation step size of 100.0 m.

The relative errors were calculated to investigate the effects caused by down-sampling. The relative error $Err_{rela}$ can be represented as (18) and (19) where $T$ and $L$ represent the time window (symbol rate×symbol number) and transmission distance, respectively. To ensure the relative error is only caused by down-sampling in frequency-domain, all the values of random parameters were generated all the same with different down-sampling ratios, such as $z_{scatt}$ in (4), $\theta$ and $\varphi$ in (9), $h_k$ and $\phi_x^{rnd}$ in (13). Based on this method, the relative error can be obtained with just single simulation, rather than repeating 10 times. Fig. 3(a) shows the relative error $Err_{rela}$ under three different walk-off conditions (0.1 ps/m, 1.0 ps/m and 10.0 ps/m). We can observe that the relative error will increase with down-sampling ratio. Further, under the same down-sampling ratio, the relative error will increase significantly with the increasing of walk-off.

$$Err_{rela}\left(R_{xt}\right) = \frac{Err_x\left(R_{xt}\right) + Err_y\left(R_{xt}\right)}{2} \quad (18)$$

$$Err_{x,y}\left(R_{xt}\right) = \frac{\int_0^T \left|E_{x,y}^B\left(L,t,R_{xt}\right) - E_{x,y}^B\left(L,t,1.0\right)\right|^2 dt}{\int_0^T \left|E_{x,y}^B\left(L,t,1.0\right)\right|^2 dt} \quad (19)$$

Further discussions about the down-sampling ratio were conducted in different situations. XT power, transmission length and calculation step size should be the three parameters quite different with those in Table 1 in a real homogeneous WC-MCF link, besides the walk-off. Here we set the down-sampling ratio to 64 to investigate the scope of validity of the channel model. As shown in Fig. 3(b) and 3(c), the relative error of XT will not change with different XT powers ranging from -55 dB/20 km to -10 dB/20 km or with transmission lengths ranging from 5.0 km to 40.0 km. Fig. 3(d) shows that the relative error will decrease with calculation step size. Therefore, it is indicated that small calculation step size and down-sampling ratio should be chosen when there is large walk-off to guarantee accuracy. The reason is large walk-off and transmission distance will introduce more serious frequency-dependent characteristics of XT, which means the sampling ratio in frequency domain should be higher or down-sampling ratio should be smaller.

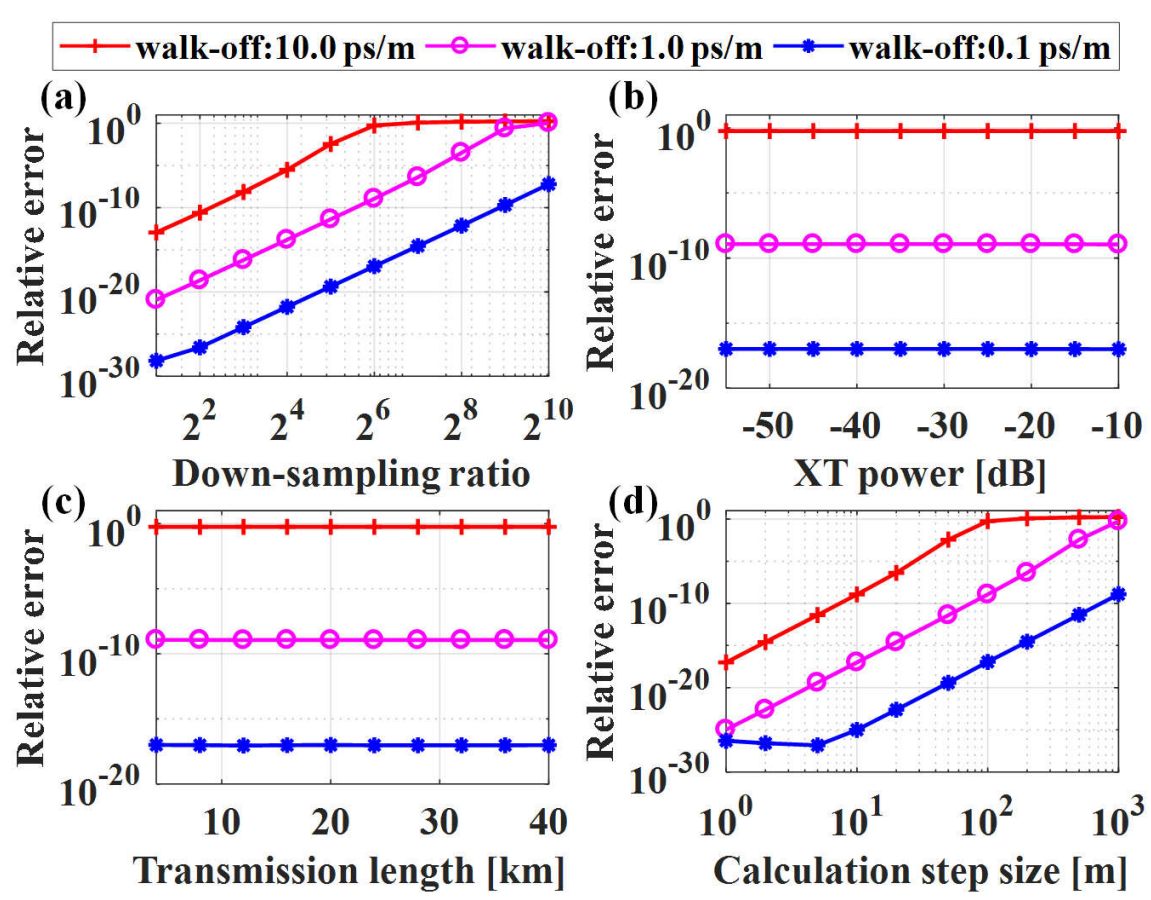

Fig. 3. Simulation error of different sweeping parameters: (a) down-sampling ratio, (b) XT power, (c) transmission distance, and (d) calculation step size.

In a real homogeneous WC-MCF, the maximum walk-off between cores is less than 0.5 ps/m in [38] and 1.3 ps/m in [9]. Therefore, it can be assumed that the largest walk-off in most of the homogeneous WC-MCF is about 1.0 ps/m. In Fig. 3(a), we can observe that the relative error is about $10^{-9}$ if the down-sampling ratio $R_{xt}$ is 64 with walk-off of 1.0 ps/m. Therefore, the down-sampling ratio $R_{xt} = 64$ should be a reasonable ratio to speed up simulation time with acceptable accuracy guaranteed for most of the real homogeneous WC-MCF.

## III. SIMULATION RESULTS OF RTD'S IMPACT ON XT

To validate whether the proposed homogeneous WC-MCF's channel model can accurately describe the frequency-dependent XT, we need to investigate the de-correlation bandwidth of XT under different conditions. From now on, it has been proved that the frequency-dependent XT is mainly caused by the frequency-dependent phase mismatch at each PMP [8], [34]. From (7), it is indicated that the frequency dependent phase mismatch could be caused by $\Delta\beta_1$, $\Delta\beta_2$, and $\Delta\beta_3$, corresponding to walk-off, the difference of chromatic dispersion, and the difference of dispersion slope, respectively, which can be represented as (20). Therefore, it is necessary to discuss the impacts of walk-off on XT.

$$\begin{aligned}\Delta\beta_1 &= \beta_1^B - \beta_1^A = 1/v_g^B - 1/v_g^A \\ \Delta\beta_2 &= \beta_2^B - \beta_2^A \\ \Delta\beta_3 &= \beta_3^B - \beta_3^A\end{aligned} \tag{20}$$

An optical pulse was injected into Core A with polarization state $x$ the same as above. The transmission distance and walk-off are 3.0 km and 0.1 ps/m, respectively, with other parameters the same as Table 1. In Core B, the output XT pulse in time-domain $XT_{time}$ after transmission can be represented as (21).

$$XT_{time}(L,t) = \left|E_x^B(L,t)\right|^2 + \left|E_y^B(L,t)\right|^2 \tag{21}$$

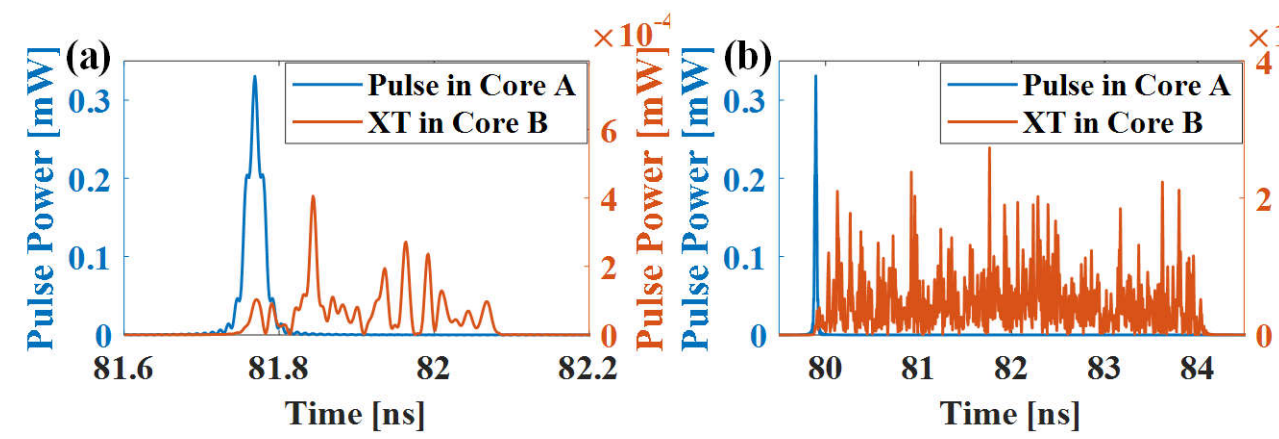

Fig. 4. (a) Simulated output pulse in two cores with walk-off of 0.1 ps/m. (b) Simulated output pulse in two cores with walk-off of 1.35 ps/m.

As shown in Fig. 4(a), the pulse width of accumulated XT in Core B after transmitted 3.0 km is about 0.3 ns. Fig. 4(b) also shows the output pulse in Core A and Core B after transmitted 3.0 km with walk-off of 1.35 ps/m. The XT's pulse width in Core B is about 4.0 ns. We can conclude that the simulation pulse widths are consistent with the target RTD (walk-off×transmission distance). In addition, due to the large RTD, the maximum XT power decreases obviously because the XT with the same energy needs to be distributed within larger time ranges.

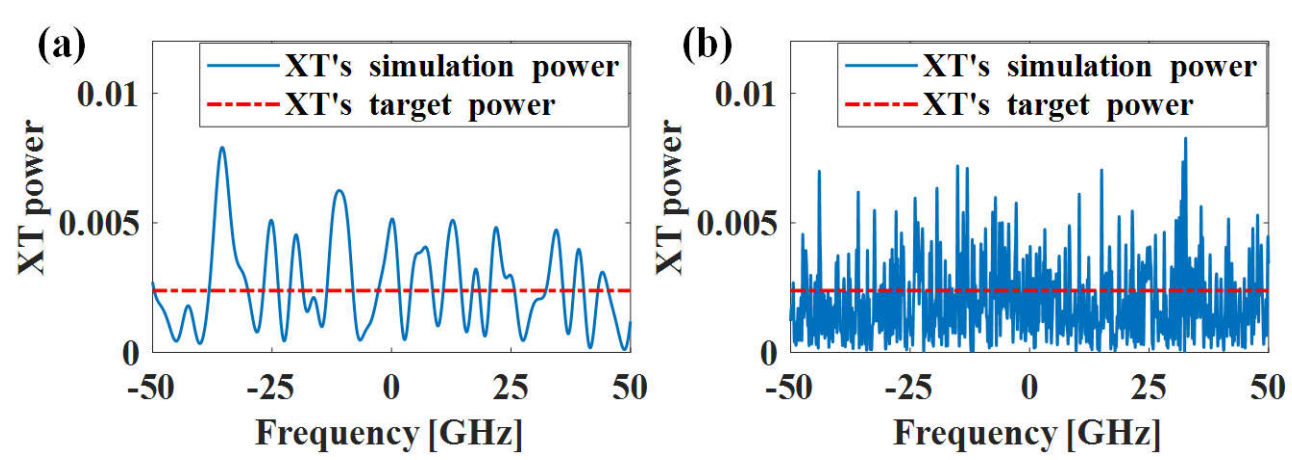

Fig. 5. (a) Simulated XT spectrum with walk-off of 0.1 ps/m. (b) Simulated XT spectrum with walk-off of 1.35 ps/m.

Fig. 5(a) and 5(b) demonstrate the corresponding XT spectrums in Fig. 4(a) and 4(b), respectively. The $XT_{fre}$ in frequency-domain can be represented as (22). As shown in Fig. 5(b), the XT with larger walk-off has more serious frequency-dependent characteristics, which indicates the XT's de-correlation bandwidth will decrease with walk-off.

$$XT_{fre}(L,\omega) = \frac{\left|E_x^B(L,\omega)\right|^2 + \left|E_y^B(L,\omega)\right|^2}{\left|E_x^A(0,\omega)\right|^2} \tag{22}$$

To figure out the relationship between XT's de-correlation bandwidth and walk-off, we sweep the walk-off from 0.1 ps/m to 1.5 ps/m with transmission distance of 3.0 km and 4.0 km. All the simulations were run 100 times to obtain the XT's statistical characteristics with other parameters the same as Table 1. The de-correlation bandwidth (half width at 1 dB) can be calculated from the autocorrelation function of XT spectrum which is represented as (22) [39]. As shown in Fig. 6, the average de-correlation bandwidth is well fitted with RTD (walk-off×transmission distance) by fractional linear function. In addition, the large fluctuation of de-correlation bandwidth under small RTD should be caused by limited frequency range ±50.0 GHz and large de-correlation bandwidth.

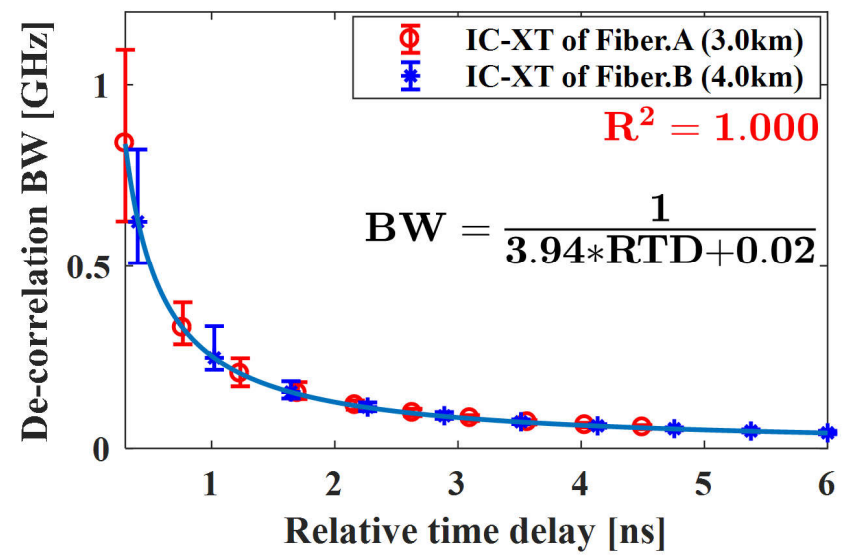

Fig. 6. Evolution of XT's de-correlation bandwidth with RTD based on the WC-MCF's channel model.

In the introduction, we mentioned that for the second strategy, it cannot truly describe the frequency-dependent XT if the XT is calculated only once in each step size when the step size is large, such as 100.0 m. To prove this statement, we can model the XT equivalently by using (11) with $h_{xt}$ of 100.0 m according to the second strategy, which means there is only one PMP within each calculation step size. The calculation step size and transmission distance are 100.0 m and 3.0 km, respectively. The walk-off is 1.35 ps/m with other parameters are the same as Table 1. Fig. 7(a) shows the simulated XT pulse. Because the XT is calculated only once in each step size, even if the XT is treated as random signals, the XT pulse still appears significant periodicity. Fig. 7(b) shows the corresponding XT spectrum. The XT's spectrum also appears significantly periodicity, which is obviously different from the XT spectrum in Fig. 5(b) and also different from the XT spectrum measured in a real homogeneous WC-MCF. These results indicate that the average occurrence distance will greatly affect the characteristics of XT itself.

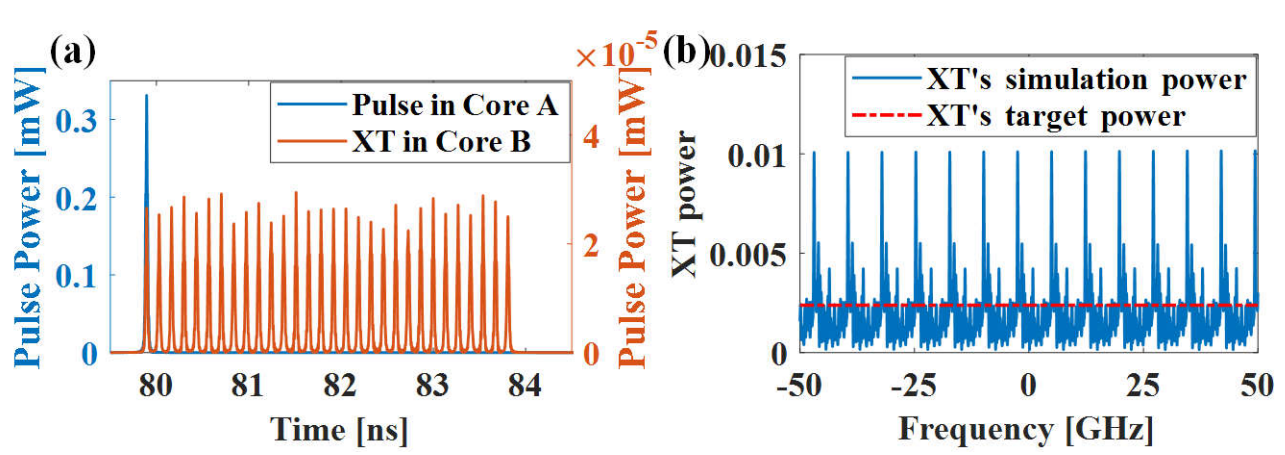

Fig. 7. Simulation using the second strategy with walk-off and calculation step size of 1.35 ps/m and 100.0 m, respectively. (a) Simulated output pulse in two cores. (b) The XT spectrum of the simulated XT output pulse.

## IV. Experimental Measurements of RTD's Impact on XT

To verify the simulation results calculated by the proposed homogeneous WC-MCF's channel model, the frequency dependent characteristics of XT are measured for our fabricated homogeneous 7-core WC-MCF. The average core pitch and cladding diameter are 41.1 μm and 150.0 μm, respectively. The XT power level is about -11.1 dB/100 km calculated by [5] assuming bending radius and twisting speed of 105.0 mm and 0.5 round/m, respectively. The measured optical parameters are shown as Table 2.

To measure the RTD between different cores, the 7-core WC-MCF was spliced with a pair of fan-in/fan-out devices [40]. One optical pulse was injected into one core, and then detected using photo diode (PD) which was being sampled by digital sampling oscilloscope (DSO) LabMaster 10Zi-A. The sample rate of DSO is 80 GSa/s, which means the measurement accuracy of time is 12.5 ps. Each core's RTD caused by fan-in/fan-out has been measured to make sure the RTD is only caused by the WC-MCF itself. Table 3 shows the measured RTD of two spools of fabricated WC-MCF with different lengths, 3.0 km (Fiber A) and 4.0 km (Fiber B). The center core (Core 1) is chosen as the reference channel. It can be observed that even if the two spools of MCF are from the same fiber preform but their RTD characteristics are different.

TABLE II
Optical Properties of Fabricated Homogeneous 7-core WC-MCF

| | Attenuation ( $dB/km$ ) | PMD ( $ps/\sqrt{km}$ ) | CD ( $ps/nm/km$ ) |
|---|---|---|---|
| Core 1 | 0.245 | 0.19 | 16.547 |
| Core 2 | 0.253 | 0.07 | 16.637 |
| Core 3 | 0.237 | 0.03 | 16.816 |
| Core 4 | 0.237 | 0.07 | 16.751 |
| Core 5 | 0.229 | 0.05 | 16.827 |
| Core 6 | 0.241 | 0.07 | 16.808 |
| Core 7 | 0.238 | 0.08 | 16.738 |

TABLE III
Measured RTD for Each Core

| | Fiber A (3.0 km) | Fiber B (4.0 km) |
|---|---|---|
| Core 1 | 0.0 | 0.0 |
| Core 2 | 1.5500 | 3.9750 |
| Core 3 | 1.8500 | 2.2875 |
| Core 4 | 1.0750 | -1.3625 |
| Core 5 | 3.7500 | 3.6625 |
| Core 6 | 4.0500 | 2.4375 |
| Core 7 | 2.1000 | 3.0125 |

* The unit of RTD is nanosecond (ns).

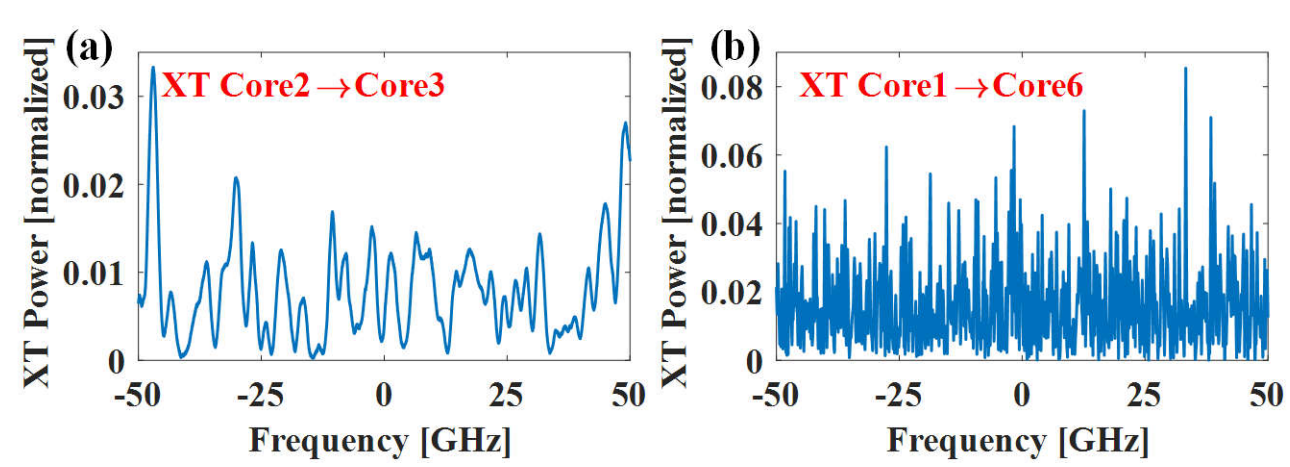

Fig. 8. (a) Measured XT spectrum with RTD of 0.3 ns (walk-off of 0.1 ps/m). (b) Measured XT spectrum with RTD of 4.05 ns (walk-off of 1.35 ps/m).

XT spectrum was also measured based on the sweeping wavelength method with optical wavelength changing from 1549.5 nm to 1550.5 nm [41]. Fig. 8(a) shows the measured XT spectrum from Core 2 to Core 3 in Fiber A with unit changing from wavelength (pm) to frequency (GHz). Fig. 8(b) shows the measured XT spectrum from Core 1 to Core 6 in Fiber A. The walk-off between Core 2 and Core 3 is 0.1 ps/m and the walk-off between Core 1 and Core 6 is 1.35 ps/m. Comparing Fig. 8 with Fig. 5, we can conclude that the simulation results are very similar to the experimental measurements.

Further, for each spool of homogeneous 7-core WC-MCF, the XT from center core (Core 1) to each outer core (Core 2-7) and the XT of each adjacent outer cores were all measured. Each inter-core XT was measured twice, such as from Core 1 to Core 2 and also from Core 2 to Core 1. As shown in Fig. 9, the measured XT's de-correlation bandwidths of two spools of WC-MCF obey the same rule even if their RTD characteristics are different. Compared with Fig. 6, we can conclude that the fitting parameters calculated by the proposed channel model are almost the same as that in a real homogeneous WC-MCF. In addition, the results indicate that the frequency-dependent XT is dominated by RTD in a real homogeneous WC-MCF.

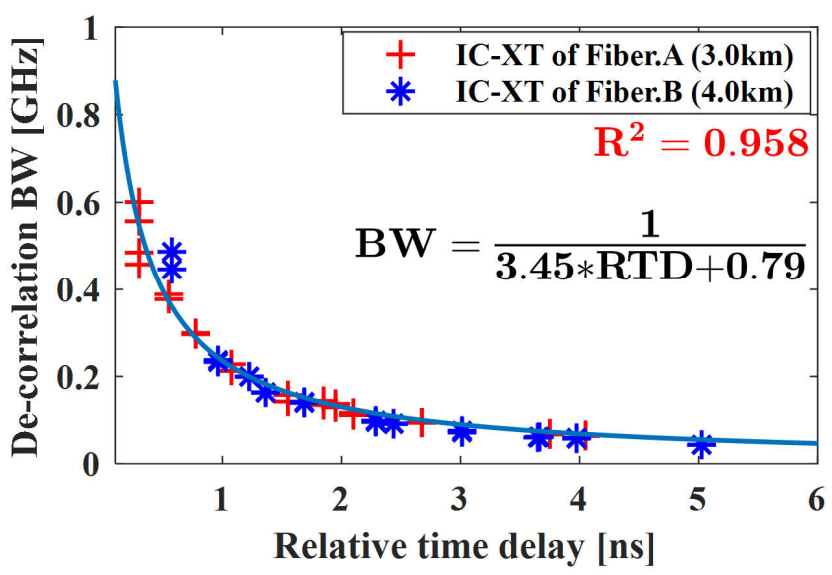

Fig. 9. Evolution of XT's de-correlation bandwidth with RTD based on the measured results.

## V. Simulation of Core Pitch's Impact on XT

When the adjacent cores get close to each other, the coupling strength between cores will grow stronger. Therefore, the signals in different cores will be fully mixed after transmission. In [14], Taiji Sakamoto et al defined SC-MCF as MCF with a Gaussian-like impulse response. To validate whether the propose channel model can accurately describe propagation effects for the homogeneous WC-MCF with different coupling strength, we swept the core pitch from 25.0 μm to 41.0 μm to obtain different coupling strength between cores. The coupling strength $K$ can be calculated based on the method introduced in [5] shown as Fig. 10. The optical center frequency, bending radius and twisting speed are 193.12 THz, 105.0 mm and 0.5 round/m, respectively. We assume the core diameter is 10.0 μm. The refractive index of core and cladding are 1.4487 and 1.4440, respectively.

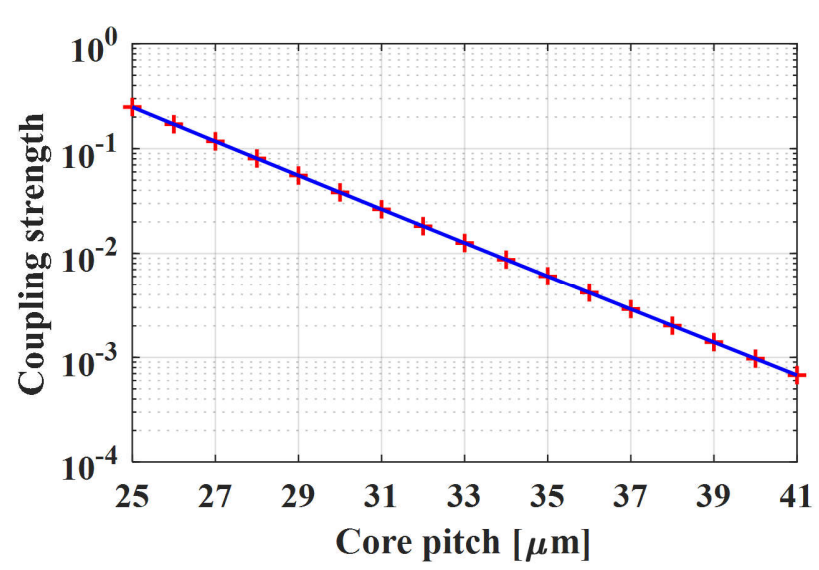

Fig. 10. Coupling strength with different core pitches.

Here, two identical optical pulses were injected into Core A and Core B simultaneously with the same polarization state $x$. The bandwidth of optical pulse is also 50.0 GHz with up-sampled by 16 times. The fiber length, calculation step size and walk-off are 3.0 km, 1.0 m and 0.1 ps/m, respectively, with other parameters the same as Table 1. Fig. 11(a) shows the output pulses of Core A and Core B with core pitch of 41.0 μm. Due to ultra-low XT power, the two pulses are separated 0.3 ns after transmission. However, when the core pitch is 28.0 μm, due to strong coupling between cores, the output pulses will not separate with each other as shown in Fig. 11(b), which has been investigated and proved in [10, 14].

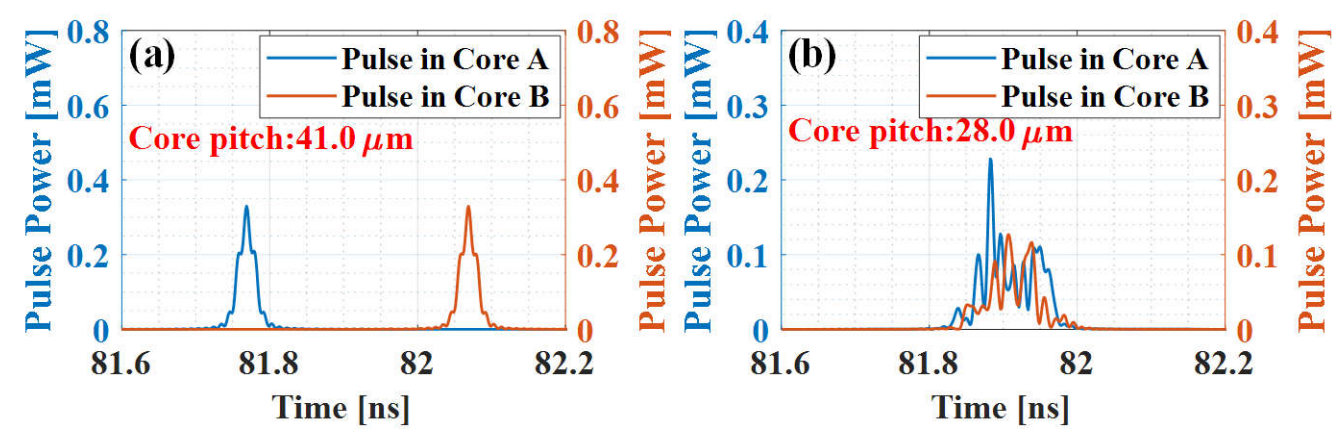

Fig. 11. (a) Simulated output pulse in two cores with core pitch 41.0 μm. (b) Simulated output pulse in two cores with core pitch 28.0 μm.

Based on the analysis of SC-MCF in [42], the pulse width (full width at 10.0 dB) will increase linearly with the square root of transmission distance if complete power mixing occurs at each PMP. But, for WC-MCF, the pulse width will increase linearly with transmission distance. Firstly, two typical core pitches were selected for analysis, which are 35.0 μm and 28.0 μm. The simulation was repeated 100 times to obtain the statistical characteristics. The walk-off, fiber length and calculation step size are 0.1 ps/m, 4.0 km and 10.0 m, respectively. The received optical pulse of direct detection can be represented as (23) to measure the impulse response.

$$I(L,t) = \left|E_x^A(L,t)\right|^2 + \left|E_y^A(L,t)\right|^2 + \left|E_x^B(L,t)\right|^2 + \left|E_y^B(L,t)\right|^2 \tag{23}$$

As shown in Fig. 12(a), the pulse width will increase linearly with transmission distance with core pitch 35.0 μm. The fitted result 0.116 ps/m is slightly larger than the walk-off of 0.1 ps/m, because CD will also spread optical pulse with 50.0 GHz bandwidth. When the core pitch is 28.0 μm, due to strong coupling between cores, the evolution of pulse width is very close to its in SC-MCF.

Further, we sweep the core pitch from 25.0 μm to 41.0 μm to obtain its relationship to the fitting parameters. Due to strong coupling caused by small core pitch, the amplitude of (12) will exceed 1.0 in one calculation step size $h$, which makes (16) cannot be solved. Therefore, it is necessary to decrease the calculation step size $h$ with small core pitch. Here, three calculation step sizes $h$ were applied, which are 1.0 m, 10.0 m and 100.0 m with simulation repeated 100 times. Fig. 12(b) shows the average fitting parameters with different core pitches. Based on the above simulation results, we can conclude that the coupling strength $K$ for WC-MCF should be less than 10-2 corresponding the core pitch about 33.0 μm. Otherwise, the MCF should not be treated as weakly-coupled, and the simplified coupling matrix (11) may be not suitable for XT analysis.

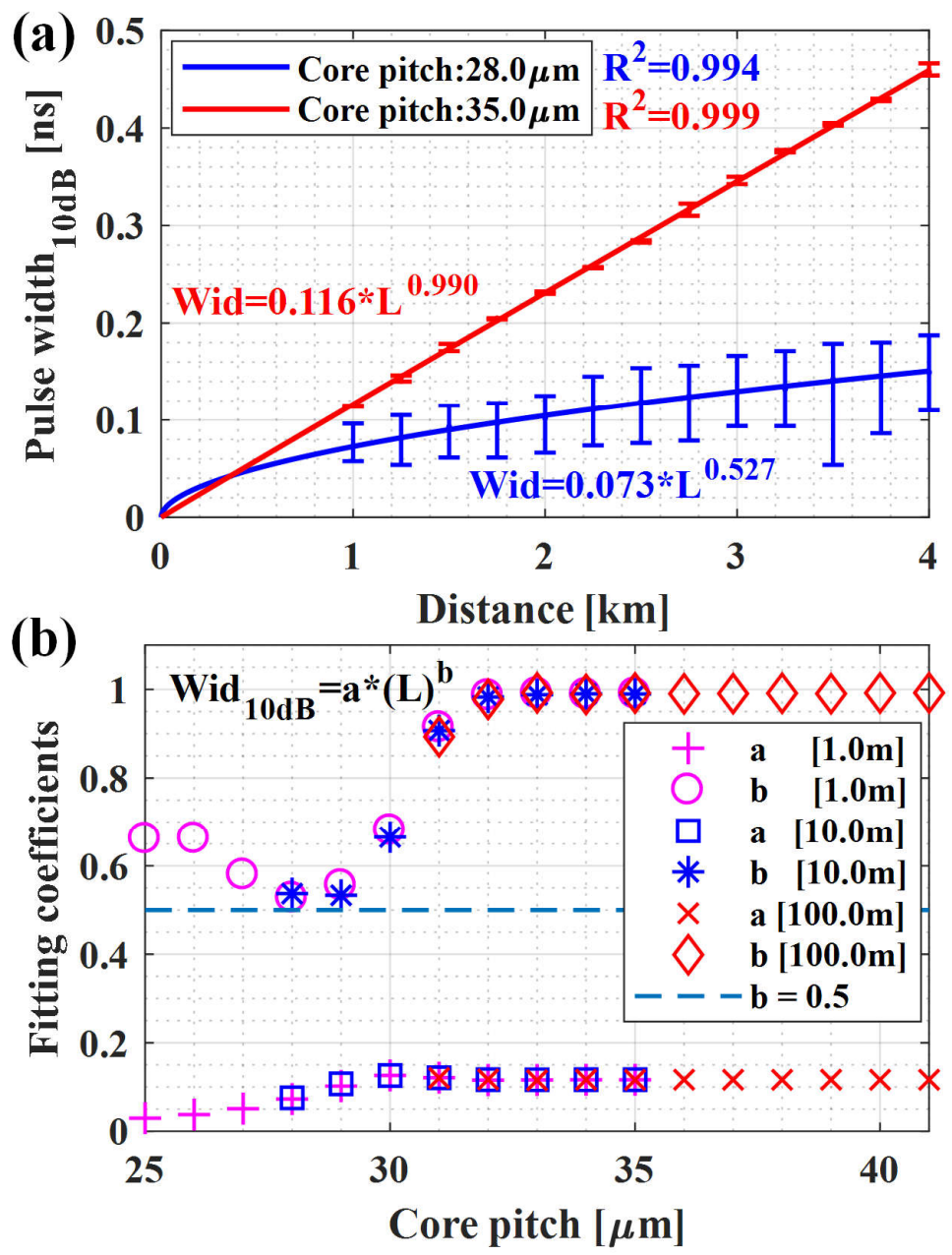

Fig. 12. (a) Fitting results of impulse response with core pitch 28.0 μm and 35.0 μm. (b) Fitting coefficients with different core pitches.

## VI. Conclusion

In this paper, we have proposed an efficient homogeneous WC-MCF's channel model based on the nonlinear Schrodinger equations. The XT between cores is modeled by a new coupling matrix. Therefore, the simulation speed can be increased with large calculation step size of 100.0 m and by down-sampling XT's coupling matrix with 64 times in frequency-domain. The down-sampling ratio 64 is reasonable when the walk-off is less than 1.0 ps/m. It can be concluded that the frequency-dependent characteristics of XT is dominated by relative time delay between cores in real homogeneous WC-MCFs. Further, we have numerically and experimentally confirmed that the de-correlation bandwidth of XT decreases with relative time delay by fractional linear function. By analyzing the characteristics of impulse response with different coupling strength, we pointed out the coupling strength should be less than $10^{-2}$ for WC-MCF. Otherwise, the XT should be more similar to it in a SC-MCF, which may be out of the scope of validity of the proposed channel model. Lastly, we believe, cooperated with the existing XT models, the proposed channel model can be used conveniently to analyze and optimize homogeneous WC-MCF based short reach systems, such as inter-data center optical interconnect, optical access network and front haul for 5G.

## Acknowledgment

L. Gan thanks Chen Yang and Weijun Tong for their support of multicore fiber, who are from Yangtze Optical Fiber and Cable Joint Stock Limited Company (YOFC), R&D center. L. Gan thanks Chen Xing and Changjian Ke for their support of crosstalk spectrum measurements. L. Gan thanks Li Shen and Liang Huo for their helpful discussions and assistance in the experiments.

## References

[1] K. Nakajima, T. Matsui, K. Saito, T. Sakamoto, and N. Araki, "Multi-Core Fiber Technology: Next Generation Optical Communication Strategy," *IEEE Communications Standards Magazine*, vol. 1, no. 3, pp. 38-45, Oct. 2017.

[2] K. Saitoh and S. Matsuo, "Multicore Fiber Technology," *J. of Lightwave Technol*, Vol. 34, no. 1, pp. 55-66, Jan. 2016.

[3] K. Saitoh and S. Matsuo, "Multicore fibers for large capacity transmission," *Nanophotonics*, vol. 2, no. 5-6, pp. 441-454, Aug. 2013.

[4] T. Hayashi, T. Nagashima, O. Shimakawa, T. Sasaki, and E. Sasaoka, "Crosstalk Variation of Multi-Core Fibre due to Fibre Bend," *in 2010 European Conference of Optical Communication (ECOC)*, Paper We.8.F.6.

[5] T. Hayashi, T. Taru, O. Shimakawa, T. Sasaki, and E. Sasaoka, "Design and fabrication of ultra-low crosstalk and low-loss multi-core fiber," *Opt. Express*, vol. 19, no. 17, pp. 16576-16592, Aug. 2011.

[6] P.J. Winzer, A.H. Gnauck, A. Konczykowska, and F. Jorge, "Penalties from in-band crosstalk for advanced optical modulation formats," in *2011 European Conference on Optical Communication (ECOC)*, Paper Tu.5.B.7.

[7] S. Yusuke, H. Keisuke, I. Itaru, M. Shoichiro, and A. Kazuhiko, "Evaluation of Inter-Core Skew in an Uncoupled Multicore Fibre," in *2017 European Conference on Optical Communication (ECOC)*, Paper W.1.B.4.

[8] L. Gan, L. Shen, M. Tang, C. Xing, Y. Li, K. Changjian, T. Weijun, L. Borui, F. Songnian, and L. Deming, "Investigation of channel model for weakly coupled multicore fiber," *Opt. Express*, vol. 26, no. 5, pp. 5182-5183, Mar. 2018.

[9] L. Gan, M. Tang, L. Shen, C. Xing, C. Ke, C. Yang, W. Tong, S. Fu, and D. Liu, "Realistic Model for Frequency-Dependent Crosstalk in Weakly-Coupled Multicore Fiber," in *2018 Optical Fiber Communication Conference (OFC)*, Paper Tu3B.6.

[10] T. Hayashi, R. Ryf, N.K. Fontaine, C. Xia, S. Randel, R. Essiambre, P.J. Winzer, and T. Sasaki, "Coupled Multicore Fiber Design with Low Inter Core Differential Mode Delay for High-Density Space Division Multiplexing," *J. of Lightwave Technol.*, vol. 33, no. 6, pp. 1175-1181, Mar. 2015.

[11] T. Hayashi, H. Chen, N.K. Fontaine, T. Nagashima, R. Ryf, R. Essiambre, and T. Taru, "Effects of Core Count/Layout and Twisting Condition on Spatial Mode Dispersion in Coupled Multi-Core Fibers," in *2016 European Conference on Optical Communication (ECOC)*, pp. 559-561.

[12] C. Xia, M.A. Eftekhar, R.A. Correa, J.E. Antonio-Lopez, A. Schulzgen, D. Christodoulides, and G. Li, "Supermodes in Coupled Multi-Core Waveguide Structures," *IEEE Journal of Selected Topics in Quantum Electronics*, vol. 22, no. 2, pp. 196-207, Mar. 2016.

[13] T. Hayashi, Y. Tamura, T. Hasegawa, and T. Taru, "Record-Low Spatial Mode Dispersion and Ultra-Low Loss Coupled Multi-Core Fiber for Ultra-Long-Haul Transmission," *J. of Lightwave Technol.*, vol. 35, no. 3, pp. 450-457, Feb. 2017.

[14] T. Sakamoto, T. Mori, M. Wada, T. Yamamoto, F. Yamamoto, and K. Nakajima, "Strongly-coupled multi-core fiber and its optical characteristics for MIMO transmission systems," *Optical Fiber Technol.*, vol. 35, pp. 8-18, Feb. 2017.

[15] R. Lin, J. Van Kerrebrouck, X. Pang, M. Verplaetse, O. Ozolins, A. Udalcovs, L. Zhang, L. Gan, M. Tang, S. Fu, R. Schatz, U. Westergren, S. Popov, D. Liu, W. Tong, T. De Keulenaer, G. Torfs, J. Bauwelinck, X. Yin, and J. Chen, "Real-time 100 Gbps/λ/core NRZ and EDB IM/DD transmission over multicore fiber for intra-datacenter communication networks," *Opt. Express*, vol. 26, no. 8, pp. 10519-10526, Apr. 2018.

[16] Z. Feng, B. Li, M. Tang, L. Gan, R. Wang, R. Lin, Z. Xu, S. Fu, L. Deng, and W. Tong, "Multicore-fiber-enabled WSDM optical access network with centralized carrier delivery and RSOA-based adaptive modulation," *IEEE Photonics Journal*, vol. 7, no. 4, pp. 1-9, Aug. 2015.

[17] Z. Feng, L. Xu, Q. Wu, M. Tang, S. Fu, W. Tong, P.P. Shum, and D. Liu, "Ultra-high capacity WDM-SDM optical access network with self-homodyne detection downstream and 32QAM-FBMC upstream," *Opt. Express*, vol. 25, no. 6, pp. 5951-5961, Mar. 2017.

[18] J.M. Galve, I. Gasulla, S. Sales, and J. Capmany, "Fronthaul design for Radio Access Networks using Multicore Fibers," *Waves*, pp. 69-80, 2015.

[19] J. He, B. Li, L. Deng, M. Tang, L. Gan, S. Fu, P.P. Shum, and D. Liu, "Experimental investigation of inter-core crosstalk tolerance of MIMO-OFDM/OQAM radio over multicore fiber system," *Opt. Express*, vol. 24, no. 12, pp. 13418-13428, Jun. 2016.

[20] T.M.F. Alves, R.S. Luís, B.J. Puttnam, A.V.T. Cartaxo, Y. Awaji, and N. Wada, "Performance of adaptive DD-OFDM multicore fiber links and its relation with intercore crosstalk," *Opt. Express*, vol. 25, no. 14, pp. 16017-16027, Jul. 2017.

[21] I.S. Chekhovskoy, V.I. Paasonen, O.V. Shtyrina, and M.P. Fedoruk, "Numerical approaches to simulation of multi-core fibers," *Journal of Computational Physics*, vol. 334, pp. 31-44, Apr. 2017.

[22] A. Macho Ortiz, C. García-Meca, F.J. Fraile-Peláez, F. Cortés-Juan, and R. Llorente Sáez, "Ultra-short pulse propagation model for multi-core fibers based on local modes," *Scientific Reports*, vol. 7, Nov. 2017.

[23] S. Mumtaz, R.J. Essiambre and G.P. Agrawal, "Nonlinear Propagation in Multimode and Multicore Fibers: Generalization of the Manakov Equations," *J. of Lightwave Technol.*, vol. 31, no. 3, pp. 398-406, Feb. 2013.

[24] C. Antonelli, M. Shtaif and A. Mecozzi, "Modeling of Nonlinear Propagation in Space-Division Multiplexed Fiber-Optic Transmission," *J. of Lightwave Technol.*, vol. 34, no. 1, pp. 36-54, Jan. 2016.

[25] C. Castro, E. De Man, K. Pulverer, S. Calabrò, M. Bohn, and W. Rosenkranz, "Simulation and Verification of a Multicore Fiber System," in *2017 19th International Conference on Transparent Optical Networks (ICTON)*, Paper We.D1.7.

[26] B. Li, L. Gan, S. Fu, Z. Xu, M. Tang, W. Tong, and P.P. Shum, "The Role of Effective Area in the Design of Weakly Coupled MCF: Optimization Guidance and OSNR Improvement," *IEEE Journal of Selected Topics in Quantum Electronics*, vol. 22, no. 2, pp. 81-87, Oct. 2015.

[27] A.V.T. Cartaxo and T.M.F. Alves, "Discrete Changes Model of Inter-core Crosstalk of Real Homogeneous Multi-core Fibers," *J. of Lightwave Technol.*, vol. 35, no. 12, pp. 2398-2408, Jun. 2017.

[28] G. Rademacher, R.S. Luís, B.J. Puttnam, Y. Awaji, and N. Wada, "Crosstalk dynamics in multi-core fibers," *Opt. Express*, vol. 25, no. 10, pp. 12020-12028, May 2017.

[29] R.O.J. Soeiro, T.M.F. Alves and A.V.T. Cartaxo, "Dual Polarization Discrete Changes Model of Inter-Core Crosstalk in Multi-Core Fibers," *IEEE Photonics Technol. Lett.*, vol. 29, no. 16, pp. 1395-1398, Jul. 2017.

[30] T.M.F. Alves and A.V.T. Cartaxo, "Characterization of the stochastic time evolution of short-term average intercore crosstalk in multicore fibers with multiple interfering cores," Opt. Express, vol. 26, no. 4, pp. 4605-4620, Feb. 2018.

[31] F. Ye, J. Tu, K. Saitoh, K. Takenaga, S. Matsuo, H. Takara, and T. Morioka, "Wavelength-Dependence of Inter-Core Crosstalk in Homogeneous Multi-Core Fibers," *IEEE Photonics Technol. Lett.*, vol. 28, no. 1, pp. 27-30, Jan. 2016.

[32] G. P. Agrawal, *Nonlinear Fiber Optics*, Academic Press, (2013).

[33] C.H. Prola, J.A.P.D. Silva, A.O.D. Forno, and R. Passy, "PMD emulators and signal distortion in 2.48-Gb/s IM-DD lightwave systems," *IEEE Photonics Technol. Lett.*, vol. 9, no. 6, pp. 842-844, Jun. 1997.

[34] A.V.T. Cartaxo, R.S. Luis, B.J. Puttnam, T. Hayashi, Y. Awaji, and N. Wada, "Dispersion Impact on the Crosstalk Amplitude Response of Homogeneous Multi-Core Fibers," *IEEE Photonics Technol. Lett.*, vol. 28, no. 17, pp. 1858-1861, Sept. 2016.

[35] A. Macho, M. Morant and R. Llorente, "Unified Model of Linear and Nonlinear Crosstalk in Multi-Core Fiber," *J. of Lightwave Technol.*, vol. 34, no. 13, pp. 3035-3046, Jul. 2016.

[36] Q. Li, *Numerical Analysis*, Tsinghua University Press, (2008).

[37] T.M.F. Alves and A.V.T. Cartaxo, "Intercore Crosstalk in Homogeneous Multicore Fibers: Theoretical Characterization of Stochastic Time Evolution," *J. of Lightwave Technol.*, vol. 35, no. 21, pp. 4613-4623, Nov. 2017.

[38] B.J. Puttnam, G. Rademacher, R.S. Luis, J. Sakaguchi, Y. Awaji, and N. Wada, "Inter-Core Skew Measurements in Temperature Controlled Multi-Core Fiber," in *2018 Optical Fiber Communication Conference*, Paper Tu3B.3.

[39] T. Hayashi, T. Sasaki and E. Sasaoka, "Behavior of Inter-Core Crosstalk as a Noise and Its Effect on Q-Factor in Multi-Core Fiber," *IEICE TRANSACTIONS ON COMMUNICATIONS*, vol. E97-B, no. 5, pp. 936-944, May 2014.

[40] B. Li, Z. Feng, M. Tang, Z. Xu, S. Fu, Q. Wu, L. Deng, W. Tong, S. Liu, and P.P. Shum, "Experimental demonstration of large capacity WSDM optical access network with multicore fibers and advanced modulation formats," *Opt. Express*, vol. 23, no. 9, pp. 10997-11006, May 2015.

[41] T.M.F. Alves and A.V.T. Cartaxo, "Characterization of Crosstalk in Ultra-Low-Crosstalk Multi-Core Fiber," *J. of Lightwave Technol.*, vol. 30, no. 4, pp. 583-589, Feb. 2012.

[42] T. Hayashi, R. Ryf, N.K. Fontaine, C. Xia, S. Randel, R. Essiambre, P.J. Winzer, and T. Sasaki, "Coupled-core multi-core fibers: High-spatial-density optical transmission fibers with low differential modal properties," in *2015 European Conference on Optical Communication (ECOC)*, Paper We.1.4.1.

**Lin Gan** was born in Hubei province, China, in 1993. He received the B.S. degree in optical information science and technology from the Huazhong University of Science and Technology (HUST), China, in 2015, and currently working toward the Ph.D. degree at Next Generation Internet Access System National Engineering Lab, HUST, China. His research interests include space division multiplexing transmission and fiber sensing.

**Jiajun Zhou** was born in Hubei province, China, in 1993. She received the B.S. degree in electronic and communication engineering from the Hubei University of Economics, China, in 2016, and currently working toward the M. Sc. degree at Next Generation Internet Access System National Engineering Lab, HUST, China. Her research interests include simplified coherent system and space division multiplexing.

**Songnian Fu** received the B.Sc. and M.Sc. degrees from Xiamen University, Xiamen, China, in 1998 and 2001, respectively, and the Ph.D. degree from Beijing Jiaotong University, Beijing, China, in 2005. From 2005 to 2011, he was with the Network Technology Research Center, Nanyang Technological University, Singapore, as a Research Fellow. Since February 2011, he has been a Professor in the School of Optical and Electronic Information and the Wuhan National Laboratory for Optoelectronics, Huazhong University of Science and Technology, Wuhan, China. His research interests include fiber-optic transmission and microwave photonics.

**Ming Tang** (SM'11) received the B.Eng. degree from Huazhong University of Science and Technology (HUST), Wuhan, China, in 2001, and the Ph.D. degree from Nanyang Technological University, Singapore, in 2005. His postdoctoral research with Network Technology Research Centre was focused on the optical fiber amplifiers, high-power fiber lasers, nonlinear fiber optics, and all-optical signal processing. From February 2009, he was with Tera-Photonics Group led by Prof. Hiromasa Ito in RIKEN, Sendai, Japan, as a Research Scientist conducting research on terahertz-wave generation, detection, and application using nonlinear optical technologies. Since March 2011, he has been a professor with the School of Optical and Electronic Information, Wuhan National Laboratory for Optoelectronics, HUST. His current research interests include optical-fiber based linear and nonlinear effects for communication and sensing applications. He has been a member of the IEEE Photonics Society since 2001 and also a member of the OSA.

**Deming Liu** received his Ph.D. degree in the department of Electronic and Information Engineering from Huazhong University of Science and Technology, (HUST), Wuhan, China, in 1999. From 1999 to 2000, he was with the Network Technology Research Center, Nanyang Technological University, Singapore, as a Research Fellow. He has been a Professor in the School of Optical and Electronic Information and the Wuhan National Laboratory for Optoelectronics, Huazhong University of Science and Technology, Wuhan, China. His research interests include optical access and wireless access.